\documentstyle[preprint,aps,epsfig]{revtex}
\begin{document}
\baselineskip=17pt
\draft
\title{Effects of neutrino mixing on high-energy cosmic neutrino flux}
\author{H. Athar$^1$, M. Je\.{z}abek$^2$ and O. Yasuda$^1$}
\address{$^1$Department of Physics, Tokyo Metropolitan University, 
 Minami-Osawa 1-1, Hachioji-Shi,\\  Tokyo 192-0397, Japan; E-mail: 
 athar@phys.metro-u.ac.jp, yasuda@phys.metro-u.ac.jp \\ 
 $^2$Institute of Nuclear Physics, Kawiory 26a, 30-055 Cracow, Poland;\\
 E-mail: marek.jezabek@ifj.edu.pl}
\date{\today}
\preprint{TMUP-HEL-0006}

\maketitle

\begin{abstract}

\tightenlines

 Several cosmologically distant astrophysical sources may produce high-energy 
 cosmic neutrinos ($E\, \geq 10^{6}$ GeV) of all flavors above the 
atmospheric neutrino background. We study the effects of vacuum neutrino 
mixing in
three flavor framework on this cosmic neutrino flux. We also consider the
effects of possible mixing between the three active neutrinos and the 
(fourth) sterile
neutrino with or without Big-Bang nucleosynthesis constraints
and estimate the resulting final high-energy cosmic neutrino flux 
ratios on earth compatible with currently existing different neutrino 
oscillation hints in a model independent way. Further, we discuss
the case where the intrinsic cosmic neutrino flux does not have the standard 
ratio.
 
\end{abstract}

\section{Introduction}

	High-energy  neutrino ($E\, \geq 10^{6}$ GeV) astroparticle physics is
now a rapidly developing field impelled by the need for improved flux 
estimates as well as a good understanding of detector capabilities for all 
neutrino flavors, particularly in light of recently growing experimental 
support for flavor oscillations \cite{skk}.

	Currently envisaged astrophysical sources of high-energy cosmic 
neutrinos include, for instance,  Active Galactic Nuclei (AGN) and Gamma Ray 
Burst fireballs (GRB)\cite{prod}. Production of high-energy cosmic neutrinos 
other than the AGNs and GRBs may also be possible \cite{td}.

	High-energy neutrino production in cosmologically distant 
astrophysical systems such as AGNs and GRBs follow mainly from the production 
and decay of relevant unstable hadrons. These unstable hadrons may be produced 
mainly when the accelerated protons in these environments interact with the 
ambient photon field ($p\gamma $) and/or protons ($pp$) present there. 
 The electron and muon 
neutrinos (and corresponding anti neutrinos) are mainly produced in the decay 
chain of charged pions whereas the tau neutrinos (and anti neutrinos) are 
mainly produced in the decay chain of charmed mesons in the same collisions 
at a suppressed level \cite{w}. 
	
	Previously the effects of vacuum neutrino mixing on the
intrinsic ratios of high-energy cosmic neutrinos, in the context of
three flavors, are briefly considered in \cite{pakvasa}. Here we
update the description \cite{paris} and consider three and four
neutrino schemes with the most up-to-date constraints from the terrestrial, 
 solar and atmospheric neutrino experiments.  We also discuss the case where
the intrinsic high-energy cosmic neutrino flux has nonstandard ratio which 
may not be obtained in charged pion decays.    

	The present study is particularly useful as the new underice/water 
\v{C}erenkov
light neutrino telescopes namely AMANDA and Baikal (also the ANTARES and 
NESTOR) as well as the large (horizontal) shower array detectors will have 
energy, angle and possibly flavor resolutions for high-energy 
neutrinos originating at cosmological distances \cite{exp}. 
 Several discussions are now
available pointing towards the possibility of flavor identification for 
high-energy cosmic neutrinos \cite{observ}.

	The plan of this paper is as follows. In section II,
we discuss the effects of vacuum neutrino mixing in three as well as
four flavor schemes and numerically estimate the final (downward going) 
 ratios of flux
for high-energy cosmic neutrinos using the ranges of neutrino mixing
parameters implied by recent searches for neutrino oscillations.
In section III, we discuss the consequences of a hypothesis in which
the intrinsic cosmic  neutrino flux does not have the standard ratio
[$F^{0}(\nu_{e})$: $F^{0}(\nu_{\mu})$:
 $F^{0}(\nu_{\tau})\, =\, 1$: $2$: 0].
In section IV, we summarize our results.

\section{Effects of neutrino mixing on high-energy cosmic neutrino flux}

	In $p\gamma$ and in $pp$ collisions, typically one obtains the 
following ratio of intrinsic high-energy cosmic neutrinos flux: 
$F^{0}(\nu_{e})$: $F^{0}(\nu_{\mu})$: $F^{0}(\nu_{\tau})\, =\, 1$: $2$:
$<10^{-5}$. For simplicity, here we take these ratios as 
 $F^{0}(\nu_{e})$: $F^{0}(\nu_{\mu})$:
 $F^{0}(\nu_{\tau})\, =\, 1$: $2$: 0. We also discuss the effects of vacuum 
neutrino mixing by varying  first two of these ratios from their above 
quoted  values as this might be the case 
under some  circumstances \cite{vary}. We count both neutrinos and anti 
neutrinos in the symbol for neutrinos. 

We consider an order of magnitude  energy 
interval essentially around $10^{6}$ GeV since the  currently available 
models for high-energy cosmic neutrinos give neutrino flux above the 
 relevant atmospheric neutrino background specifically 
from AGN cores within this energy interval. Also in this 
energy range the flavor identification for high-energy cosmic neutrinos may 
be conceivable in new km$^{2}$ surface area neutrino telescopes \cite{observ}.
We have checked that for high-energy neutrinos originating from 
cosmologically distant sources, the change in the flavor composition of the
high-energy cosmic neutrinos due to vacuum mixing is essentially energy 
independent for the entire energy range relevant for observations  
 as the energy effects are  averaged out in the relevant neutrino
flavor oscillation probability expressions.
 
		It has been pointed out that there are no matter effects 
on vacuum neutrino oscillations for relevant mass squared difference values 
 [${\cal O} (10^{-10})\leq \Delta m^{2}/\mbox{eV}^{2}\leq {\cal O} (1)$] for 
 high-energy cosmic neutrinos scattering over the matter particles in the 
vicinity of sources \cite{athar}. Nevertheless, for high-energy cosmic 
neutrinos scattering over the relic neutrinos, in halos or other wise, during 
their propagation, there may be relatively small (at most, of the order of few 
percent) matter enhanced flavor oscillation effects under the assumption of 
 rather strong 
CP asymmetric neutrino background \cite{ls}. However, given the 
current status of high-energy cosmic neutrino flux measurements, vacuum flavor
oscillations remain the dominant effect and therefore from now on we consider 
here the effects of vacuum flavor 
oscillations only. For some other possible mixing effects, see \cite{npb}. 
The typical distance to these astrophysical sources is 
taken as $L\, \simeq \, 10^{2}$ Mpc (where 1 pc $\simeq \, 3\cdot 10^{18}$ cm).
It is relevant here to mention that our following analysis is not necessarily 
restricted
to high-energy cosmic neutrinos originating only from AGNs or GRBs as the above
mentioned ratios of intrinsic neutrino flux can in principle be conceivable 
from  some other cosmologically distant astrophysical sources as 
well \cite{td}. For simplicity, we assume the absence of relatively dense 
objects between the cosmologically distant high-energy neutrino sources and 
the prospective neutrino telescopes so as not to change significantly the
vacuum oscillation pattern.   
	
\subsection{Three neutrino scheme}

It has been known in the two flavor framework that the solar and the
atmospheric neutrino problems are accounted for by neutrino
oscillations with $\Delta m^2_{\rm
smaller}\sim 10^{-5}$eV$^2$ or $10^{-10}$eV$^2$ and $\Delta m^2_{\rm
larger}\sim 10^{-2.5}$eV$^2$, respectively.  Without loss of
generality we assume that $|\Delta m_{21}^2|<|\Delta m_{32}^2|<|\Delta
m_{31}^2|$ where $\Delta m^2_{ij}\equiv m^2_i-m^2_j$ ($i\neq j$; $i, j = 
 1,2,3$).  If both the
solar and the atmospheric neutrino problems are to be
solved by energy dependent solutions, we have to have $\Delta
m_{21}^2\simeq\Delta m^2_\odot$ and $\Delta m_{32}^2\simeq\Delta
m^2_{\mbox{\rm atm}}$, i.e., we have mass hierarchy in this case.
 Here, the subscripts $\odot$ and atm stand for the mixing angles
for the solar and atmospheric neutrino oscillations, respectively.

In the three flavor framework, the flavor and mass eigenstates are 
related by the MNS matrix U \cite{mns}:
\begin{eqnarray}
\left( \begin{array}{c} \nu_e  \\ \nu_{\mu} \\ 
\nu_{\tau} \end{array} \right)
=U\left( \begin{array}{c} \nu_1  \\ \nu_2 \\ 
\nu_3 \end{array} \right),\quad
U&\equiv&\left(
\begin{array}{ccc}
U_{e1} & U_{e2} &  U_{e3}\\
U_{\mu 1} & U_{\mu 2} & U_{\mu 3} \\
U_{\tau 1} & U_{\tau 2} & U_{\tau 3}
\end{array}\right).
\label{eqn:mns}
\end{eqnarray}
In the present convention for the mass pattern,
the disappearance probability for the CHOOZ experiment \cite{chooz} is given by
\begin{eqnarray}
P(\bar{\nu}_e\rightarrow\bar{\nu}_e)=1-4|U_{e3}|^2(1-|U_{e3}|^2)
\sin^2\left({\Delta m_{32}^2L \over 4E}\right),
\label{eqn:chooz}
\end{eqnarray}
and combining the negative result by the CHOOZ experiment\cite{chooz} with
the atmospheric neutrino data, we have (for quantitative discussions, 
 see \cite{y1}):
\begin{eqnarray}
|U_{e3}|^2\ll 1.
\end{eqnarray}

As was mentioned in the Introduction, the matter effect is irrelevant
in our discussions, so let us consider the vacuum oscillation probability in
the case where the oscillation length is very small as compared to the 
distance between the source and the detector [hereafter referred to as far 
 distance (approximation)].  In vacuum, it is well
known that the flavor oscillation probability is given by
\begin{eqnarray}
P(\nu_\alpha\rightarrow\nu_\beta;L)=\delta_{\alpha\beta}
-\sum_{j\ne k}U_{\alpha j}^\ast U_{\beta j}
U_{\alpha k} U_{\beta k}^\ast\left(
1-e^{-i\Delta E_{jk}L}\right).
\label{eqn:prob}
\end{eqnarray}
In the limit $L\rightarrow\infty$, we have
\begin{eqnarray}
P(\nu_\alpha\rightarrow\nu_\beta;L=\infty)=\delta_{\alpha\beta}
-\sum_{j\ne k}U_{\alpha j}^\ast U_{\beta j}
U_{\alpha k} U_{\beta k}^\ast
&=&\sum_{j}|U_{\alpha j}|^2 |U_{\beta j}|^2,
\end{eqnarray}
where we have averaged over rapid oscillations.
Thus, we can represent the oscillation probability as a symmetric matrix $P$
and $P$ can be written as a product of a matrix $A$:
\begin{eqnarray}
P\equiv\left(
\begin{array}{ccc}
P_{ee} & P_{e\mu} &  P_{e\tau}\\
P_{e\mu} & P_{\mu\mu} & P_{\mu\tau} \\
P_{e\tau} & P_{\mu\tau} & P_{\tau\tau}
\end{array}\right)\equiv AA^T,
\end{eqnarray}
with
\begin{eqnarray}
A\equiv\left(
\begin{array}{ccc}
|U_{e1}|^2 & |U_{e2}|^2 &  |U_{e3}|^2\\
|U_{\mu 1}|^2 & |U_{\mu 2}|^2 & |U_{\mu 3}|^2 \\
|U_{\tau 1}|^2 & |U_{\tau 2}|^2 & |U_{\tau 3}|^2
\end{array}\right).
\end{eqnarray}
Now, the cosmic neutrino flux in the far distance can be expressed as a product
of $P$ and the intrinsic flux $F^0(\nu_\alpha) (\alpha=e,\mu,\tau)$:
\begin{eqnarray}
\left( \begin{array}{c} F(\nu_e)  \\ F(\nu_\mu) \\ 
F(\nu_\tau) \end{array} \right)
=P\left( \begin{array}{c} F^0(\nu_e)  \\ F^0(\nu_\mu) \\ 
F^0(\nu_\tau) \end{array} \right)
=AA^T\left( \begin{array}{c} F^0(\nu_e) \\ F^0(\nu_\mu) \\ 
F^0(\nu_\tau) \end{array} \right).
\end{eqnarray}
Throughout this section we assume the standard ratio of
the intrinsic cosmic  neutrino flux, i.e., 
$F^{0}(\nu_{e})$: $F^{0}(\nu_{\mu})$:
 $F^{0}(\nu_{\tau})\, =\, 1$: $2$: 0, so that, we get
\begin{eqnarray}
A^T\left( \begin{array}{c} F^0(\nu_e) \\ F^0(\nu_\mu) \\ F^0(\nu_\tau)
\end{array} \right) &=&\left( \begin{array}{ccc} |U_{e1}|^2 & |U_{\mu
1}|^2 & |U_{\tau 1}|^2\\ |U_{e2}|^2 & |U_{\mu 2}|^2 & |U_{\tau 2}|^2\\
|U_{e3}|^2 & |U_{\mu 3}|^2 & |U_{\tau 3}|^2 \end{array}\right) \left(
\begin{array}{c} 1 \\ 2 \\ 0 \end{array} \right)F^0(\nu_e), \nonumber\\
&=&\left(
\begin{array}{c} 1 \\ 1 \\ 1 \end{array} \right)F^0(\nu_e) +\left(
\begin{array}{c} |U_{\mu 1}|^2-|U_{\tau 1}|^2 \\ |U_{\mu
2}|^2-|U_{\tau 2}|^2 \\ |U_{\mu 3}|^2-|U_{\tau 3}|^2 \end{array}
\right)F^0(\nu_e),
\label{eqn:flux}
\end{eqnarray}
where we have used
the unitarity condition, i.e.,  $\sum_j |U_{\alpha j}|^2=$1.  
 When $|U_{e3}|^2\ll 1$ and $|U_{\mu 3}|\simeq
|U_{\tau 3}|$, we have $||U_{\mu j}|^2-|U_{\tau j}|^2|\ll 1
(j=1,2,3)$, so the second term in Eq. (\ref{eqn:flux}) is small.  Hence
with the constraints of the solar and atmospheric neutrino and the reactor
data, we obtain
\begin{eqnarray}
\left( \begin{array}{c} F(\nu_e)  \\ F(\nu_\mu) \\ 
F(\nu_\tau) \end{array} \right)
\simeq
A\left( \begin{array}{c} 1  \\1 \\ 
1\end{array} \right)F^0(\nu_e)
\simeq
\left( \begin{array}{c} 1  \\1 \\ 
1\end{array} \right)F^0(\nu_e),
\end{eqnarray}
where we have used the unitarity condition again.  Therefore, we
conclude that the ratio of the cosmic high-energy neutrino fluxes in
the far distance is 1: 1: 1, irrespective of which solar solution is
chosen.  This is because the MNS matrix elements satisfy
$|U_{e3}|^2\ll 1$, $|U_{\mu 3}|\simeq |U_{\tau 3}|$ and
$F^{0}(\nu_{e})$: $F^{0}(\nu_{\mu})$: $F^{0}(\nu_{\tau})\, =\, 1$:
$2$: 0.  This is the feature that is expected to be observed at the
new km$^2$ surface area neutrino telescopes in the case of the three
neutrino oscillation scheme.

Using the allowed region for the MNS matrix elements given in \cite{y1},
we have evaluated the final ratio of the cosmic neutrino flux numerically.
To plot the ratio of the three neutrino flavors, we introduce the triangle
representation, which is introduced by Fogli, Lisi, and Scioscia \cite{fls}
in a different context.  In Fig. 1, a unit regular triangle is drawn
and the position of the point gives the ratio of the final high-energy 
neutrino flux with $F_\alpha\equiv F(\nu_\alpha)$.
One reason that we adopt this representation is because currently 
we do not know the precise total cosmic neutrino flux, while we may  
observe the ratio of flux of different cosmic  neutrino flavors 
 experimentally
to a certain extent.  Using this triangle representation, the allowed
region in the three flavor framework with the constraints from the
terrestrial, solar
and atmospheric neutrino data, is given in Figs. 2a and 2b.  The allowed
region is a small area around the midpoint $F(\nu_e)=F(\nu_\mu)=
F(\nu_\tau)=1/3$ and
the small deviation from the midpoint indicates the smallness of
$|U_{e3}|$ and $||U_{\mu j}|^2-|U_{\tau j}|^2|$.

\subsection{Four neutrino scheme} 

Let us now consider the case with three active and one sterile
neutrino.  Schemes with sterile neutrinos have been studied by a
number of authors \cite{sterile,goswami,oy,bggs}.  Here, we are interested in
the four neutrino scheme in which the solar, atmospheric and LSND experiment 
 \cite{lsnd} are all explained by neutrino oscillations.  By
generalizing the discussion on Big-Bang Nucleosynthesis (BBN) from the
two neutrino scheme \cite{bbn} to four neutrino case, it has been
shown \cite{oy,bggs} that the neutrino mixing angles are strongly constrained
not only by the reactor data \cite{goswami} but also by BBN if one
demands that the number $N_\nu$ of effective neutrinos be less than
four.  In this case, the $4\times 4$ MNS matrix splits approximately
into two $2\times 2$ block diagonal matrices.

On the other hand, some authors have given conservative estimate for
$N_\nu$ \cite{he4} and if their estimate is correct then we no longer
have strong constraints on the neutrino mixing angles.  In this section, we 
 discuss
the final ratio of the cosmic neutrino flux components with and without
BBN constraints separately.  Throughout this section we assume the following 
mass pattern
\begin{eqnarray}
  m_1^2\lesssim  m_2^2\ll  m_3^2\lesssim  m_4^2,
\end{eqnarray}
with corresponding $\Delta m_{21}^2$, $\Delta m_{43}^2$ and
$\Delta m_{32}^2$ stand for the mass squared differences suggested by the
solar, atmospheric neutrino deficits and the LSND experiment,
respectively.

\subsubsection{Four neutrino scheme with BBN constraints}

Here, we adopt the notation in \cite{oy} for the $4\times 4$ MNS matrix:
\begin{eqnarray}
U&\equiv&\left(
\begin{array}{cccc}
U_{e1} & U_{e2} &  U_{e3} &  U_{e4}\\
U_{\mu 1} & U_{\mu 2} & U_{\mu 3} & U_{\mu 4}\\
U_{\tau 1} & U_{\tau 2} & U_{\tau 3} & U_{\tau 4}\\
U_{s1} & U_{s2} &  U_{s3} &  U_{s4}
\end{array}\right), \nonumber\\
&\equiv&R_{34}({\pi \over 2}-\theta_{34})R_{24}(\theta_{24})
R_{23}({\pi \over 2})
e^{2i\delta_1\lambda_3}R_{23}(\theta_{23})
e^{-2i\delta_1\lambda_3}
e^{\sqrt{6}i\delta_3\lambda_{15}/2}\nonumber\\
&\times&R_{14}(\theta_{14})e^{-\sqrt{6}i\delta_3\lambda_{15}/2}
e^{2i\delta_2\lambda_8/\sqrt{3}}R_{13}(\theta_{13})
e^{-2i\delta_2\lambda_8/\sqrt{3}}
R_{12}(\theta_{12}),
\label{eqn:u}
\end{eqnarray}
where $c_{ij}\equiv\cos\theta_{ij}$, $s_{ij}\equiv\sin\theta_{ij}$ and
\begin{eqnarray}
R_{jk}(\theta)\equiv \exp\left(iT_{jk}\theta\right),
\end{eqnarray}
is a $4\times4$ orthogonal matrix with
\begin{eqnarray}
\left(T_{jk}\right)_{\ell m}=i\left(\delta_{j\ell}\delta_{km}
-\delta_{jm}\delta_{k\ell}\right),
\end{eqnarray}
and $2\lambda_3\equiv{\rm diag}(1,-1,0,0)$,
$2\sqrt{3}\lambda_8\equiv{\rm diag}(1,1,-2,0)$,
$2\sqrt{6}\lambda_{15}\equiv{\rm diag}(1,1,1,-3)$
are diagonal elements of the $SU(4)$ generators.
From the constraint of the reactor data of Bugey \cite{bugey},
which strongly constrain the disappearance probability
$P(\bar{\nu}_e\rightarrow\bar{\nu}_e)$ for the entire region
of mass squared difference implied by the data of LSND \cite{lsnd}
and E776 \cite{e776}, we have \cite{goswami,oy,bggs}
\begin{eqnarray}
|U_{e3}|^2+|U_{e4}|^2\ll 1.
\end{eqnarray}
On the other hand, from the BBN constraint that sterile neutrino
be not in thermal equilibrium, we get
\begin{eqnarray}
|U_{s3}|^2+|U_{s4}|^2\ll 1.
\end{eqnarray}
In this case, the oscillation probabilities in the far distance are
given essentially by the following two neutrino flavor formulae:
\begin{eqnarray}
P(\nu_e\rightarrow\nu_e;L=\infty)&=&
1-{1 \over 2}|U_{e1}|^2|U_{e2}|^2=1-{1 \over 2}\sin^22\theta_{12}, \nonumber\\
P(\nu_e\rightarrow\nu_s;L=\infty)&=&
{1 \over 2}|U_{e1}|^2|U_{e2}|^2={1 \over 2}\sin^22\theta_{12}, \nonumber\\
P(\nu_\mu\rightarrow\nu_\mu;L=\infty)&=&
1-{1 \over 2}|U_{\mu3}|^2|U_{\mu4}|^2=
1-{1 \over 2}\sin^22\theta_{24}, \nonumber\\
P(\nu_\mu\rightarrow\nu_\tau;L=\infty)&=&
{1 \over 2}|U_{\mu3}|^2|U_{\mu4}|^2={1 \over 2}\sin^22\theta_{24},
\end{eqnarray}
where, we have substituted  $\theta_{13}=\theta_{14}=\theta_{23}=\theta_{34}=0$
in Eq. (\ref{eqn:u}) and $\theta_{12}$, $\theta_{24}$ correspond to
the mixing angles of the solar and atmospheric neutrino oscillations,
respectively.  It has been known \cite{solar} that only the Small
Mixing Angle (SMA) MSW solution is allowed in the
$\nu_e\leftrightarrow\nu_s$ solar oscillation scheme, so in the
present case, we take $|\theta_{12}|\ll 1$.  Thus, we have the
following ratio of the final cosmic high-energy neutrino flux:
\begin{eqnarray}
\left( \begin{array}{c} F(\nu_e)  \\ F(\nu_\mu) \\ 
F(\nu_\tau) \\F(\nu_s)  \end{array} \right)
&=&
\left( \begin{array}{cccc} 1 & 0& 0&0\\
0&0&c_{\mbox{\rm atm}}^2&s_{\mbox{\rm atm}}^2\\
0&0&s_{\mbox{\rm atm}}^2&c_{\mbox{\rm atm}}^2\\
0&1&0&0\end{array}\right)
\left( \begin{array}{cccc} 1 & 0& 0&0\\
0&0&0&1\\
0&c_{\mbox{\rm atm}}^2&s_{\mbox{\rm atm}}^2&0\\
0&s_{\mbox{\rm atm}}^2&c_{\mbox{\rm atm}}^2&0\end{array}\right)
\left( \begin{array}{c} 1\\ 2\\ 0\\0
\end{array} \right)F^0(\nu_e), \nonumber\\
&=&\left( \begin{array}{c} 1\\ 2-\sin^22\theta_{\mbox{\rm atm}}\\ 
\sin^22\theta_{\mbox{\rm atm}}\\0
\end{array} \right) F^0(\nu_e),
\end{eqnarray}
where, we have taken $\theta_{24}\equiv\theta_{\mbox{\rm atm}}$ which
satisfies \cite{toshito}
\begin{eqnarray}
0.88\lesssim \sin^22\theta_{\mbox{\rm atm}} \le 1.
\label{eqn:atmsin}
\end{eqnarray}
In this case, therefore, we again have
$F(\nu_{e})$: $F(\nu_{\mu})$:
 $F(\nu_{\tau})\, \simeq\, 1$: $1$: 1.

We have also numerically obtained the allowed region by letting
$\theta_{12}$, $\theta_{13}$, $\theta_{14}$, $\theta_{23}$,
$\theta_{34}$ and $\theta_{24}$ to vary within the constraints obtained from 
 the reactor, solar and atmospheric neutrino data.  In the four neutrino
scheme, the total flux of active neutrino is in general not the same as
that at the production point, and in principle we have to use a unit
tetrahedron to express the ratio of the four neutrino flux.  In
practice, however, we may not observe the cosmic sterile neutrino events nor
do we know the precise total flux of the cosmic high-energy neutrinos,
so it is useful to normalize the flux
of each active neutrino by the total flux of active neutrinos:
\begin{eqnarray}
\left( \begin{array}{c} \widetilde{F}(\nu_e)  \\ \widetilde{F}(\nu_\mu) \\ 
\widetilde{F}(\nu_\tau) \end{array} \right)
\equiv{1 \over F(\nu_e)+F(\nu_\mu)+F(\nu_\tau)}
\left( \begin{array}{c} F(\nu_e)  \\ F(\nu_\mu) \\ 
F(\nu_\tau) \end{array} \right).
\end{eqnarray}
After redefining the flux, we can plot the ratio of each active neutrino
with the same triangle graph as in the three neutrino case, and
the result is depicted in Fig. 3.  The allowed region is again
small and the reason that the region lies horizontally is because
of possible deviation of $\theta_{24}\equiv\theta_{\mbox{\rm atm}}$
from $\pi/4$ [see Eq. (\ref{eqn:atmsin})].

\subsubsection{Four neutrino scheme without BBN constraints}

In this subsection, we discuss what happens to the ratio of the final cosmic
high-energy neutrino flux if we lift BBN constraints.  Without BBN
constraints, the only conditions we have to take into account are the
solar and atmospheric neutrino deficit data and the results of experiment of 
LSND
(the appearance experiment of $\bar{\nu}_\mu\rightarrow\bar{\nu}_e$) 
 \cite{lsnd},
Bugey (the disappearance experiment of
$\bar{\nu}_e\rightarrow\bar{\nu}_e$) \cite{bugey} and CDHSW (the
disappearance experiment of $\bar{\nu}_\mu\rightarrow\bar{\nu}_\mu$)
 \cite{cdhsw}.
For the range of the $\Delta m^{2}$
suggested by the LSND data, which is given by 0.2$\lesssim 
\Delta m^{2}_{\mbox{\rm{\scriptsize LSND}}}/\mbox{eV}^{2}\lesssim$ 2  when
combined with the data of Bugey \cite{bugey} and E776 \cite{e776},
the constraint by the Bugey data is very stringent and
\begin{eqnarray}
|U_{e3}|^2+|U_{e4}|^2\lesssim 10^{-2},
\end{eqnarray}
has to be satisfied \cite{goswami,oy,bggs}. Therefore, for simplicity, we take
$U_{e3}=U_{e4}=0$, in the following discussion.

The analysis of the solar neutrino data in the four neutrino scheme
with ansatz $U_{e3}=U_{e4}=0$ has been done recently by Giunti,
Gonzalez-Garcia and Pe\~na-Garay \cite{ggp}.  They have shown that the
scheme is reduced to the two neutrino framework in which only one free
parameter $c_s\equiv|U_{s1}|^2+|U_{s2}|^2$ 
appears in the
analysis\footnote{In the notation of \cite{ggp} this parameter is
given by $c_s=c_{23}^2 c_{24}^2$. We adopt different notation
from \cite{ggp} for the parametrization of the mixing angles, however,   
will use $c_s$ for $|U_{s1}|^2+|U_{s2}|^2$.}.  Their conclusion
is that the SMA (MSW) solution exists for the entire region of $0\le c_s
\le 1$, while the Large Mixing Angle (LMA) and Vacuum Oscillation (VO)
solutions survive only for $0\le c_s \lesssim 0.2$ and $0\le c_s \lesssim
0.4$, respectively.

The analysis of the atmospheric neutrino data in the four neutrino
framework has been done by one of the authors \cite{y2} more recently
again with ansatz $U_{e3}=U_{e4}=0$.  The conclusion of \cite{y2} is
that the region $-\pi/2 \lesssim\theta_{34} \lesssim \pi/2$ and
$0 \lesssim \theta_{23} \lesssim \pi/6$ as well as 
  $0\le \delta_1\le \pi$ is consistent
with the Superkamiokande atmospheric neutrino data of the contained
and upward going $\mu$ events, where $\theta_{34}$ and $\theta_{23}$
stand for, roughly speaking, the ratio of
$\nu_\mu\leftrightarrow\nu_\tau$ versus $\nu_\mu\leftrightarrow\nu_s$ 
and the ratio of the contributions of
$\sin^2\left(\Delta m^{2}_{\mbox{\rm atm}}L/4E\right)$
versus
$\sin^2\left(\Delta m^2_{\mbox{\rm{\scriptsize LSND}}}L/4E\right)$
in the flavor oscillation probability, respectively, and $\delta_1$ is the only
 CP phase left in this scheme.  Notice that the recent claim by
the Superkamiokande group \cite{toshito} that $\nu_\mu\leftrightarrow\nu_s$
is almost completely excluded is based on the hypothesis of  two
neutrino $\nu_\mu\leftrightarrow\nu_s$ oscillations with
only one mass squared difference $\Delta m^{2}_{\mbox{\rm atm}}$,
and their claim is completely consistent with the results
in \cite{y2}, where the region $\theta_{34} \simeq \pm\pi/2$
and $\theta_{23}\simeq 0$, which would lead to pure
$\nu_\mu\leftrightarrow\nu_s$ oscillations, is excluded.
For generic mixing angles of $\theta_{34}$ and $\theta_{23}$, 
the ansatz in \cite{y2} implies hybrid of
$\nu_\mu\leftrightarrow\nu_\tau$ and $\nu_\mu\leftrightarrow\nu_s$ 
oscillations in general, and one has to take into account the constraint
of the CDHSW experiment \cite{cdhsw} also.  Following \cite{y2},
we take $\Delta m^{2}_{32}\equiv\Delta m^{2}_{\mbox{\rm{\scriptsize LSND}}}=$ 0.3 eV$^2$, 
which is consistent with the negative result of the disappearance experiment
of CDHSW \cite{cdhsw} for the entire region of the mixing angles
obtained in \cite{y2}.

Combining the results of \cite{ggp} on solar neutrinos and \cite{y2} on
atmospheric neutrinos, we have evaluated the final ratio of the cosmic 
 high-energy 
neutrino flux and the result is given in Figs. 4a and 4b,
where the allowed region with the SMA (MSW), LMA (MSW) and VO solutions are
shown separately.  Almost the entire allowed region with the SMA (MSW)
solution lies above the line $\widetilde{F}(\nu_e)=1/3$, and
this is because the total normalization $F(\nu_e)+
F(\nu_\mu)+F(\nu_\tau)=1-F(\nu_s)$ becomes less than 1 while
$F(\nu_e)=(1-\sin^22\theta_{12})F^{0}(\nu_e)$ hardly changes due to
smallness of $|\theta_{12}|=|\theta_\odot|$.
On the other hand, for most of the region with LMA (MSW) and VO solutions,
$F(\nu_e)/F^{0}(\nu_e)$ becomes smaller than $1-F(\nu_s)$ and the region
lies below the line $\widetilde{F}(\nu_e)=1/3$.
This four neutrino scheme without BBN constraint  gives us clearly a 
distinctive pattern for the final ratio of the cosmic high-energy neutrino 
flux, and observationally, if we have good precision then it may be even 
possible to distinguish
the SMA (MSW) solution from the LMA (MSW) or VO solutions.
The allowed regions in Figs. 4a and 4b are plotted for
$\delta_1=0$ and $\delta_1=\pi/2$.  We observe that most
of the allowed regions for $\delta_1=0$ and $\delta_1=\pi/2$
overlap with each other and it implies that distinction
between $\delta_1=0$ and $\delta_1=\pi/2$ is difficult in this scheme of 
four neutrinos.

We note in passing that the allowed region of the
LOW solution to the solar netrino problem is contained in
that of the LMA solution as far as $\sin^22\theta_\odot$ is
concerned, and therefore the LOW solution gives us
only a subset of the allowed region of the LMA solution
in Figs. 4a and 4b.

\section{Nonstandard ratio of the high-energy neutrino flux} 

In section II, we have
seen that the schemes of three neutrinos and of four neutrinos with
BBN constraints give us the ratio $F(\nu_{e})$: $F(\nu_{\mu})$:
$F(\nu_{\tau})\, \simeq\, 1$: $1$: 1, irrespective of which solar
solution is chosen.  This is due to the fact that the intrinsic high-energy 
 cosmic neutrino flux has
the ratio $F^{0}(\nu_{e})$: $F^{0}(\nu_{\mu})$: $F^{0}(\nu_{\tau})\,
=\, 1$: $2$: 0 and the oscillation of atmospheric neutrinos is with maximal
mixing while $|U_{e3}|^2\ll 1$.  It has been pointed out \cite{vary}
that the intrinsic flux of the cosmic high-energy neutrinos may not have
the standard ratio $F^{0}(\nu_{e})$: $F^{0}(\nu_{\mu})$:
$F^{0}(\nu_{\tau})\, =\, 1$: $2$: 0, mainly because some of muons may lose
their energy in a magnetic field.  Here, we discuss in a model
independent way the consequences of a generic scenario which is
characterized by $F^{0}(\nu_{e})$: $F^{0}(\nu_{\mu})$:
$F^{0}(\nu_{\tau})\, =\, \lambda /3$: $1-\lambda /3$: 0, where $\lambda $ is
a free parameter $0\le \lambda \le 1$ (the standard ratio is obtained for
$\lambda=1$).

We have plotted the allowed region in Figs. 5a and 5b for $\lambda=1/2$ and
$\lambda = 0$, respectively.  We observe that the cosmic high-energy neutrino
 flux with
the nonstandard ratio gives relatively lower value of $\widetilde{F}(\nu_e)$.
If relatively lower value of cosmic $\nu_e$ flux is observed in the future 
 experiments,
then it may imply that the possible oscillation scenario is 
the four neutrino scheme without BBN constraints and
the intrinsic cosmic neutrino flux with the nonstandard ratio.
 Independent information on neutrino mixing parameters as well as on relevant
astrophysical inputs may be needed here to arrive at a more definite 
conclusion.

\section{Results and Discussion}

	In general, the final flux of high-energy cosmic neutrinos is expected
to be almost equally distributed among the three (active) cosmic neutrino 
flavors because of vacuum flavor mixing/oscillations provided the astrophysical
sources for these high-energy cosmic neutrinos are cosmologically distant, 
 essentially 
 irrespective of the neutrino flavors (three or four). Nevetheless, 
 this may not be the case if the intrinsic high-energy cosmic neutrino flux 
ratios differ from the standard one, namely from $F^{0}(\nu_{e})$: 
 $F^{0}(\nu_{\mu})$: $F^{0}(\nu_{\tau})\, =\, 1$: $2$: 0. In the examples 
 considered in this work for the nonstandard intrinsic high-energy cosmic 
neutrino flux, a relatively lower final flux for cosmic $\nu_{e}$ is obtained.
 The situation of nonstandard intrinsic cosmic neutrino flux may arise, for 
instance, if some of the muons lose their energy in the relatively intense 
magnetic field in the vicinity of the source.

	A simultaneous measurement of the 
three cosmic neutrino flavors may be useful to obtain information 
about a particular neutrino (mass and) mixing scheme depending on the relevant 
achievable resolutions for typical km$^{2}$ surface area neutrino telescopes.

 Irrespective of the numbers of neutrino flavors, in each of the
neutrino (mass and) mixing scheme discussed in this work, the final
flux of high-energy cosmic tau neutrinos is essentially comparable to
that of non tau (active) neutrinos, even if it is intrinsically negligible.
This may, at least in principle, be useful to constrain the relevant 
nonstandard particle physics/astro physics scenarios. 

	In this work, we have considered the effects of vacuum neutrino flavor
oscillations on high-energy cosmic neutrino flux in the context of three as 
well as four flavors. These oscillations result in an energy independent ratio,
$R_{\alpha \beta}\equiv N_{\alpha}/N_{\beta}\, \,  (\alpha \neq \beta; \alpha 
 = \beta
 = \nu_{e}, \nu_{\mu}, \nu_{\tau})$ of the number of events detected for the
neutrino flavors $\alpha $ and $\beta $. It is so because the various neutrino
flavor precession probabilities given in section II [see Eq. (5) and Eq. (17)]
 are essentially 
independent of neutrino energy. In the following paragraph, we briefly 
describe the prospects offered by the typical km$^{2}$ surface area under 
ice/water neutrino telescopes which are currently under construction/planning 
to possibly identify the cosmic neutrino flavor and hence to determine the 
 ratio, $R_{\alpha \beta}$ \cite{observ}.

	We ignore the possible observational difference between cosmic 
neutrinos and anti neutrinos for simplicity in the following discussion and 
assume that the flavor content in the cosmic neutrino flux is equally 
 distributed 
because of vacuum flavor oscillations. Several of the recent discussions 
suggest that the absorption of high-energy cosmic neutrino flux by earth is 
neutrino flavor dependent \cite{observ}. The upward going electron and muon 
neutrino fluxes are
significantly attenuated typically for $E_{0}\geq 5\cdot 10^{4}$ GeV, 
whereas the upward going tau neutrinos with $E > E_{0} $ may reach the 
detector with $E \leq  E_{0}$ because of the short life time of the associated 
tau lepton and may appear as a rather small pile up with fairly flat zenith 
angle  dependence. For $E \geq  E_{0}$, the upward going cosmic neutrino event
 rates range typically as: $N_{\nu_{\mu}}\sim 
{\cal O}(10^{1})$ whereas $ N_{\nu_{\tau}}\sim {\cal O}(10^{0})$ in units of
per year per steradian for typical km$^{2}$ surface area neutrino telescopes,
if one uses the current upper high-energy cosmic neutrino flux limits 
\cite{prod}. Let us
note that for $E \geq E_{0}$, presently the high-enegy cosmic neutrino fluxes 
from AGNs dominate above the atmospheric neutrino background. For downward 
going high-enegy cosmic neutrino flux, the 
event rate ranges typically as: 
 $N_{\nu_{e}}\sim {\cal O}(10^{1.5}), N_{\nu_{\mu}}
\sim {\cal O}(10^{2})$, whereas $ N_{\nu_{\tau}}\sim {\cal O}(10^{1})$ in 
units of per year per steradian for the same high-energy cosmic neutrino 
fluxes. For $E > E_{0}$, the downward going cosmic tau neutrinos typically 
produce a two bang event topology such that the two bangs are connected by a 
$\mu$-like track. The size of the second bang being on the average a factor of
two larger than the first bang. The downward going electron neutrinos produce 
a single bang at these energies whereas the muon neutrinos typically produce 
a single shower alongwith a zipping $\mu$-like track in km$^{2}$ surface area
neutrino telescopes. Based on this rather distinct event topologies, cosmic 
neutrino flavor identification may be conceivable. The above order of 
 magnitude estimates indicate that the typical 
km$^{2}$ surface area neutrino telescopes do offer some prospects for 
observations of high-energy cosmic neutrino flavor ratio, $R_{\alpha \beta}$,
 or at least may constrain it meaningfully.\\

\paragraph*{Acknowledgments}

	The work of H. A. is supported by a Japan Society for the Promotion
of Science fellowship. The authors thank the hospitality of 
 Summer Institute 99 at Yamanashi, Japan where this work was
started.  This research was supported in
part by a Grant-in-Aid for Scientific Research of the Ministry of
Education, Science and Culture, Japan (\#12047222, \#10640280).

\newpage

\noindent
{\Large{\bf Figures}}

\begin{description}
\item[Fig.1] The representation of the ratio of the final flux of 
 (downward going) high-energy  cosmic neutrinos on earth  
 with a unit regular triangle.  The $x$ and $y$ coordinates
 of the points are given by $x=(2F_\tau+F_e)/\sqrt{3}$, $y=F_e$ and
 the inside region of the triangle is given by 
 $0\le x\le 2/\sqrt{3}$, $0\le y \le 1$. The point with
 an asterisk stands for the ratio without oscillations
 $F(\nu_{e})$: $F(\nu_{\mu})$: $F(\nu_{\tau})\, =\, 1$: $2$: 0 
 and is given by the coordinate
 $(1/3\sqrt{3}, 1/3)$.

\item[Fig.2] The ratio of the final flux of high-energy cosmic neutrinos
 in the far distance in the three neutrino scheme.
 The allowed region is the inside of each contour;
 (b) is an enlarged figure of (a).  The allowed region lies near the mid point
 $(1/\sqrt{3}, 1/3)$.

\item[Fig.3] The ratio of the final flux of high-energy cosmic neutrinos
 in the far distance in the four neutrino scheme with BBN
 constraints.  The allowed region lies near the mid point
 $(1/\sqrt{3}, 1/3)$.

\item[Fig.4] The ratio of the final flux of high-energy cosmic neutrinos
 in the far distance in the four neutrino scheme without BBN
 constraints;  (b) is an enlarged figure of (a).
The allowed region of the LOW solution is a subset of that of
the LMA solution.

\item[Fig.5] The ratio of the final flux of high-energy cosmic neutrinos
 in the far distance in a nonstandard scenario  characterized
 by $F^{0}(\nu_{e})$: $F^{0}(\nu_{\mu})$: $F^{0}(\nu_{\tau})
 \, =\, \lambda /3$: $1-\lambda/3$: 0.
 In Fig. 5(a), $\lambda =1/2$, the cases of $N_\nu=3$ with SMA and
 $N_\nu=4$ with BBN constraints have small region near the point
 $F(\nu_{e})$: $F(\nu_{\mu})$: $F(\nu_{\tau})
 \, =\,$ 2: 5: 5, the region for the cases of
 $N_\nu=3$ with LMA and VO lie above this point, whereas most of the
region
 for the case of $N_\nu=4$ without BBN constraints lie to the left of this
 point. As in Fig.4, the allowed region of the LOW solution is a
subset of that of the LMA solution.
In Fig. 5(b), $\lambda =0$, the cases of $N_\nu=3$ with SMA and
 $N_\nu=4$ with BBN constraints have small region near the point
 $F(\nu_{e})$: $F(\nu_{\mu})$: $F(\nu_{\tau})
 \, =\, 0$: $1$: 1, the region for the cases of
 $N_\nu=3$ with LMA and VO lie above this point, whereas most of the region 
 for the case of $N_\nu=4$ without BBN constraints lie to the left of this
 point.
\end{description}

\newpage
\vglue 3cm
\epsfig{file=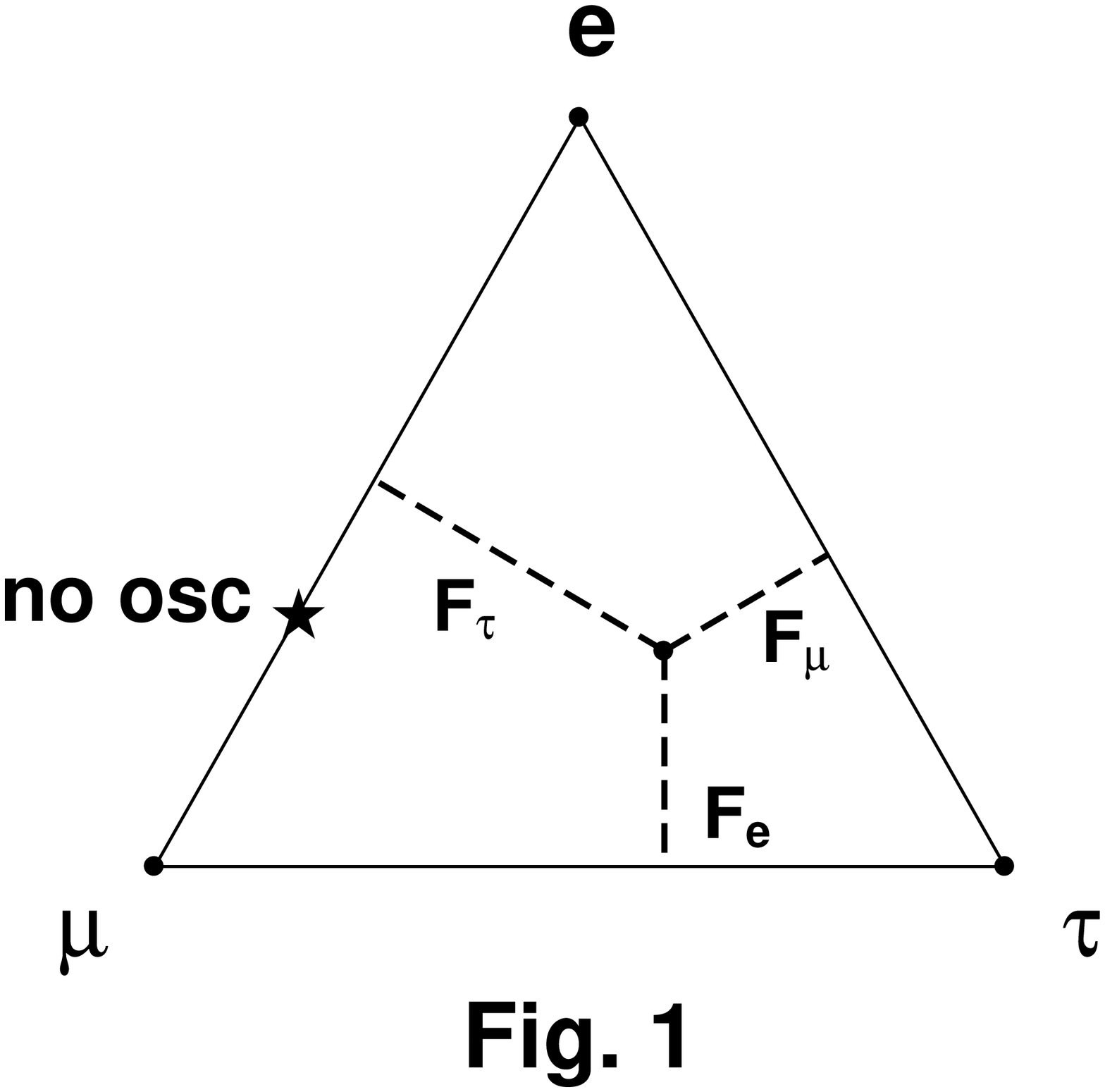,width=15cm}
\newpage
\vglue -14.3cm
\hglue 0.2cm\epsfig{file=fig2a.eps,width=12cm}
\vglue -3cm
\hglue 0.7cm\epsfig{file=fig2b.eps,width=12cm}
\newpage
\vglue 8cm
\epsfig{file=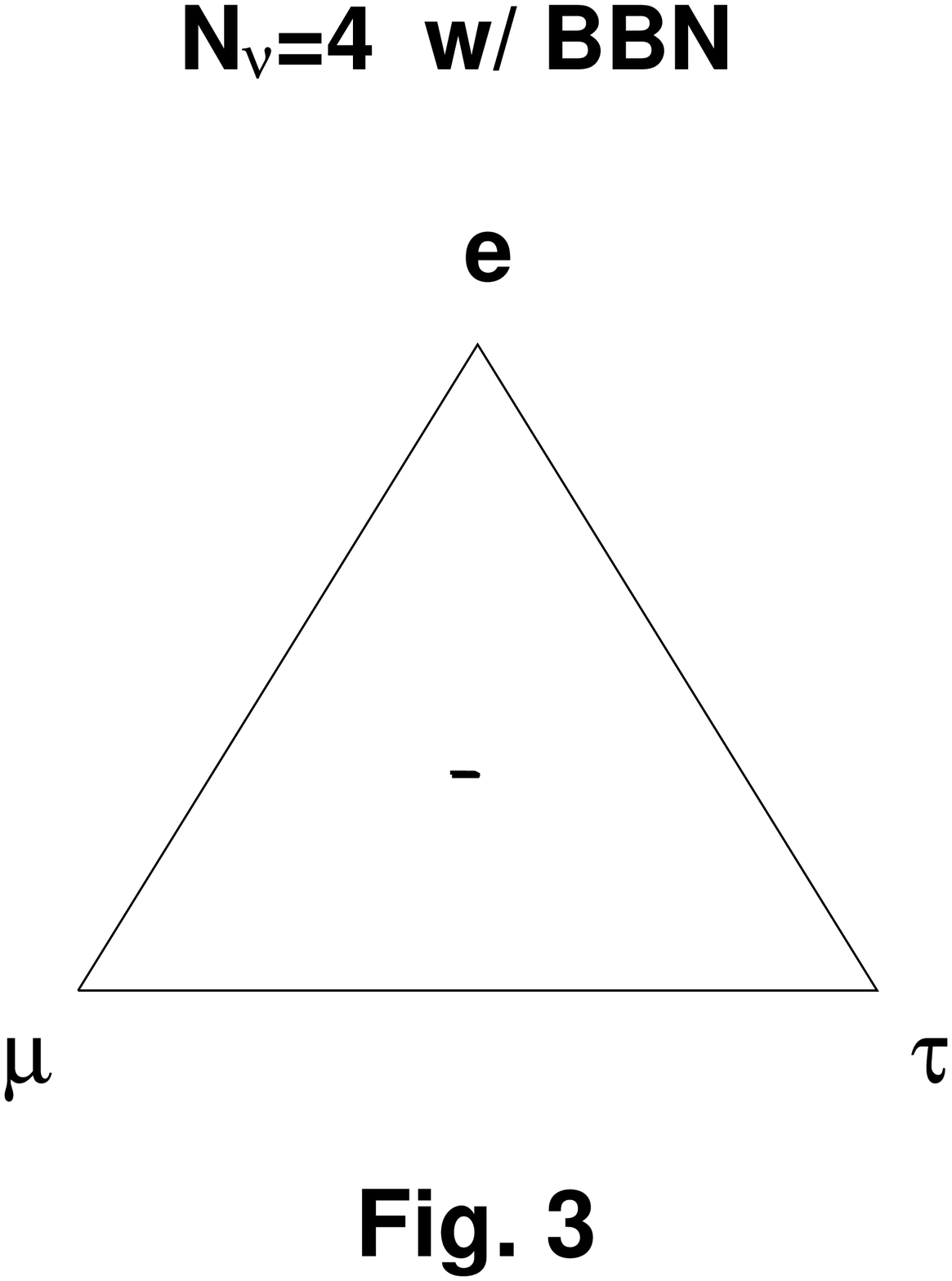,width=15cm}
\newpage
\vglue -14.3cm
\hglue 2cm\epsfig{file=fig4a.eps,width=12cm}
\vglue 2cm
\hglue 2.5cm\epsfig{file=fig4b.eps,width=12cm}
\newpage
\vglue -14.3cm
\hglue 2cm\epsfig{file=fig5a.eps,width=12cm}
\vglue 2cm
\hglue 2.5cm\epsfig{file=fig5b.eps,width=12cm}

\end{document}